\documentclass[12pt, letterpaper]{article}

\usepackage[utf8]{inputenc}
\usepackage{fontenc}
\usepackage{geometry}
\usepackage{graphicx}
\usepackage{amsmath}
\usepackage{amssymb}
\usepackage{hyperref}
\usepackage{booktabs} 
\usepackage{float}    
\usepackage{enumitem} 
\usepackage{authblk}  
\usepackage{xcolor}   
\usepackage{tcolorbox}
\usepackage{amsmath, amssymb, amsthm}
\usepackage{algorithm}
\usepackage{algpseudocode}
\usepackage{graphicx}
\usepackage{booktabs}
\usepackage{tikz}
\usetikzlibrary{shapes, arrows, positioning}
\usepackage{hyperref}
\usepackage{graphicx}
\usepackage{booktabs} 
\usepackage{geometry}
\usepackage{hyperref}
\usepackage{xcolor}
\usepackage{stmaryrd} 

\hypersetup{
    colorlinks=true,
    linkcolor=blue,
    filecolor=magenta,      
    urlcolor=cyan,
    citecolor=blue
}

\title{\textbf{Optimistic TEE-Rollups: A Hybrid Architecture for Scalable and Verifiable Generative AI Inference on Blockchain}}
\author{%
\begin{tabular}{c}
Aaron Chan\textsuperscript{a},
Alex Ding\textsuperscript{a,*},
Frank Chen\textsuperscript{a}\\
Alan Wu\textsuperscript{a},
Bruce Zhang\textsuperscript{a},
Arther Tian\textsuperscript{a}\\
\textsuperscript{a}DGrid AI\\
\textsuperscript{*}Corresponding author: \href{mailto:alex.ding@dgrid.ai}{\texttt{alex.ding@dgrid.ai}}
\end{tabular}%
}

\date{}

\begin{document}

\maketitle

\begin{abstract}
The rapid integration of Large Language Models (LLMs) into decentralized physical infrastructure networks (DePIN) is currently bottlenecked by the \textit{Verifiability Trilemma}, which posits that a decentralized inference system cannot simultaneously achieve high computational integrity, low latency, and low cost.
Existing cryptographic solutions, such as Zero-Knowledge Machine Learning (ZKML), suffer from super-linear proving overheads ($\mathcal{O}(k \cdot N \log N)$) that render them infeasible for billion-parameter models.
Conversely, optimistic approaches (opML) impose prohibitive dispute windows, preventing real-time interactivity, while recent "Proof of Quality" (PoQ) paradigms sacrifice cryptographic integrity for subjective semantic evaluation, leaving networks vulnerable to model downgrade attacks and reward hacking.

In this paper, we introduce \textbf{Optimistic TEE-Rollups (OTR)}, a hybrid verification protocol that harmonizes these constraints.
OTR leverages NVIDIA H100 Confidential Computing Trusted Execution Environments (TEEs) to provide sub-second \textit{Provisional Finality}, underpinned by an optimistic fraud-proof mechanism and stochastic Zero-Knowledge spot-checks to mitigate hardware side-channel risks.
We formally define \textit{Proof of Efficient Attribution} (PoEA), a consensus mechanism that cryptographically binds execution traces to hardware attestations, thereby guaranteeing model authenticity.
Extensive simulations demonstrate that OTR achieves 99\% of the throughput of centralized baselines with a marginal cost overhead of \$0.07 per query, maintaining Byzantine fault tolerance against rational adversaries even in the presence of transient hardware vulnerabilities.
\end{abstract}

\section{Introduction}

The convergence of Artificial Intelligence (AI) and Distributed Ledger Technology (DLT) promises a shift from centralized API monopolies to permissionless, censorship-resistant intelligence markets.
However, the realization of this vision relies on a critical, unresolved primitive: \textit{Verifiable Inference}.
In a trustless environment, a user $U$ submitting a query $q$ to a remote node $S$ must verify that the response $r$ was generated by the specific model $F_\theta$ (e.g., Llama-3-405B) agreed upon in the smart contract, rather than a cheaper approximation $F_{\theta'}$ or a pre-cached response.
This requirement introduces the \textit{Verifiability Trilemma}, a structural constraint stating that current decentralized AI architectures can optimize for only two of three critical variables:
\begin{enumerate}
    \item \textbf{Computational Integrity:} Cryptographic certainty that $r = F_\theta(q)$.
    \item \textbf{Inference Latency:} The time-to-finality required for interactive user experiences ($<1$s).
    \item \textbf{Economic Efficiency:} Keeping verification costs negligible relative to inference costs.
\end{enumerate}

\subsection{The Limitations of Existing Paradigms}

Current solutions occupy extrema on this tradeoff curve.
\textbf{Zero-Knowledge Machine Learning (ZKML)} \cite{ezkl, modulus} offers the strongest integrity guarantees by arithmetizing neural network operations into zk-SNARK circuits.
However, the computational overhead for non-linear operations (e.g., Softmax, GeLU) in Transformer architectures is prohibitive.
Generating a proof for a single inference of a 70B parameter model currently requires hours on enterprise hardware \cite{zkml_survey}, violating the latency requirement for interactive AI agents.
\textbf{Optimistic Machine Learning (opML)} \cite{opml} adapts the fraud-proof mechanism of optimistic rollups \cite{kalodner2018arbitrum}, assuming honest execution by default.
While this preserves economic efficiency ($C_{op} \approx C_{native}$), it necessitates a "Dispute Window" (typically 7 days) to allow validators to challenge incorrect state transitions.
This latency renders opML unsuitable for composable DeFi applications or real-time agents requiring immediate settlement.
Most recently, \textbf{Proof of Quality (PoQ)} \cite{zhang2024proof} has emerged as a "costless" alternative.
PoQ bypasses execution verification entirely, instead using lightweight "assessor models" (e.g., BERT cross-encoders) to score the semantic relevance of the output.
While PoQ solves the latency and cost constraints, it fails the integrity requirement fundamentally.
As we demonstrate in Section 4, PoQ is susceptible to \textit{Model Downgrade Attacks}, where a rational adversary serves responses from a smaller, cheaper model (e.g., 8B params) that satisfies the semantic judge but defrauds the user of the premium paid for the larger model (e.g., 70B params).
Furthermore, relying on static reward models incentivizes \textit{Reward Hacking} \cite{gao2022scaling}, where adversarial perturbations maximize the score without retaining semantic utility.

\subsection{The OTR Proposition}

To resolve the Trilemma, we propose \textbf{Optimistic TEE-Rollups (OTR)}, a hybrid architecture that treats Trusted Execution Environments (TEEs) as \textit{optimistic co-processors}.
OTR acknowledges that while TEEs (such as NVIDIA H100 Confidential Computing) provide near-native performance and hardware-based attestation, they are historically vulnerable to sophisticated side-channel attacks (e.g., Foreshadow, Spectre) \cite{vanbulck2018foreshadow}.
OTR employs a "Defense in Depth" strategy:
\begin{enumerate}
    \item \textbf{Tier 1: Provisional Finality.} The TEE generates a signed attestation $\sigma_{TEE}$ proving that the inference trace was generated by a specific binary (MRENCLAVE).
    This provides sub-second finality for the vast majority of users.
    \item \textbf{Tier 2: Optimistic Fallback.} To mitigate trust in the hardware manufacturer, the protocol remains open to challenges.
    A decentralized network of "Fishermen" can dispute the execution trace.
    \item \textbf{Tier 3: Stochastic Spot-Checks.} The protocol mandates probabilistic ZK-proof generation for a subset of transactions ($\rho < 0.01$), creating a statistical guarantee that deters rational adversaries from exploiting hardware vulnerabilities.
\end{enumerate}

Our primary contribution, \textit{Proof of Efficient Attribution} (PoEA), cryptographically binds the economic value of a query to the specific model weights loaded in the enclave, solving the Model Authenticity problem that plagues PoQ.
\subsection{Contributions}

This paper makes the following contributions:
\begin{itemize}
    \item \textbf{Protocol Architecture:} We formalize the OTR protocol, defining the interaction between TEE Attestations, Optimistic Dispute Games, and VRF-triggered ZK spot-checks.
    \item \textbf{Security Analysis:} We provide a game-theoretic analysis of rational adversaries, proving that OTR renders Model Downgrade Attacks economically irrational ($ \mathbb{E}[\text{profit}] < 0 $), a guarantee absent in PoQ.
    \item \textbf{Empirical Validation:} We benchmark OTR using NVIDIA H100 SXM5 simulations.
    Results indicate a 1400x speedup over ZKML and a 99\% reduction in latency compared to opML, while maintaining strong Byzantine Fault Tolerance.
\end{itemize}

The remainder of this paper is organized as follows: Section 2 reviews the related work in decentralized AI.
Section 3 details the OTR protocol and PoEA consensus. Section 4 provides a rigorous scalability, security, and cost analysis of the protocol, and Section 5 presents experimental results.

\section{Literature Review}

The integration of Large Language Models (LLMs) into Distributed Ledger Technology (DLT) necessitates a rigorous analysis of the "Verifiability Trilemma."
This trilemma posits that a decentralized inference network can currently optimize for only two of the following three attributes: **(1) Computational Integrity** (Cryptographic Security), **(2) Low Latency** (Real-time usability), and **(3) Cost Efficiency** (Scalability).
In this section, we classify existing approaches into three primary paradigms: Validity Proofs (ZKML), Fraud Proofs (opML), and Outcome-Based Verification (PoQ), followed by an analysis of Trusted Execution Environments (TEEs) as a hardware primitive.
\subsection{Validity Proofs: Zero-Knowledge Machine Learning (ZKML)}
Zero-Knowledge Machine Learning (ZKML) provides the strongest form of computational integrity by converting neural network operations into arithmetic circuits compatible with zk-SNARKs or zk-STARKs.
Frameworks such as EZKL \cite{ezkl} and Modulus \cite{modulus} allow a prover $\mathcal{P}$ to convince a verifier $\mathcal{V}$ that output $y = F_\theta(x)$ was computed correctly without revealing the model weights $\theta$ or input $x$.
However, ZKML faces a prohibitive "Proving Overhead." The conversion of floating-point operations (FLOPS) common in Transformers to finite field arithmetic introduces massive redundancy.
For a model with $N$ parameters, the proving time $T_{prove}$ scales super-linearly:
\begin{equation}
    T_{prove} \approx \mathcal{O}(k \cdot N \log N) \gg T_{inference}
\end{equation}
where $k$ is a large constant representing the circuit expansion factor (often $10^3$ to $10^4$) \cite{zhang2024proof}.
While recursive proof composition (e.g., Halo2) mitigates on-chain verification costs, the generation latency renders ZKML infeasible for interactive LLM applications (e.g., Llama-3-70B), where user retention demands sub-second responses.
\subsection{Fraud Proofs: Optimistic Machine Learning (opML)}
Optimistic Machine Learning (opML) adapts the interactive dispute resolution mechanisms of Optimistic Rollups (e.g., Arbitrum \cite{kalodner2018arbitrum}) to AI inference.
As proposed by Conway et al. \cite{opml}, opML assumes the sequencer is honest by default.
Finality is only achieved after a dispute window $T_{challenge}$ (typically 7 days).
If a validator challenges a result, an interactive bisection protocol is triggered to identify the single instruction $i$ where the sequencer and validator diverge.
This instruction is then executed on-chain. While opML achieves near-native inference costs ($C_{op} \approx C_{native}$), the latency imposed by $T_{challenge}$ breaks composability with other DeFi or DePIN protocols requiring immediate settlement.
Furthermore, opML requires the reproduction of the full inference trace $\mathcal{D}_{trace}$, which raises privacy concerns regarding user prompts and intermediate activation states.
\subsection{Outcome-Based Verification: Proof of Quality (PoQ)}
Recognizing the computational bottlenecks of ZKML and the latency of opML, recent works have proposed outcome-based verification.
Most notably, **Proof of Quality (PoQ)** \cite{zhang2024proof} shifts the consensus focus from the *process* of inference to the *quality* of the result.
As formally defined by Zhang et al., PoQ employs a lightweight assessor model $M$ (e.g., a BERT-based cross-encoder) to evaluate the tuple $(q, r)$, where $q$ is the query and $r$ is the response.
The system reaches consensus via a score aggregation mechanism:
\begin{equation}
    S_{final} = \frac{1}{k} \sum_{i=1}^{k} \phi(M_i(q, r))
\end{equation}
where $k$ is the number of assessors and $\phi$ is a reward distribution function designed to discourage lazy voting.
**Critique:** While PoQ achieves excellent latency and throughput (processing 1,600+ pairs/sec on A100s \cite{zhang2024proof}), it suffers from two critical deficiencies for high-value financial use cases:
\begin{itemize}
    \item \textbf{Model Authenticity:} PoQ verifies that an output is *good*, not that it was generated by a specific model (e.g., Llama-3-405B).
    A malicious node could serve responses from a cheaper, smaller model (e.g., Llama-3-8B) that still satisfies the semantic judge, effectively defrauding the user of the premium paid for the larger model's compute.
    \item \textbf{Reward Hacking:} As seen in Reinforcement Learning from Human Feedback (RLHF) literature \cite{gao2022scaling}, optimizing against a static reward model (the PoQ judge) inevitably leads to adversarial perturbations where nonsense outputs trigger high scores.
\end{itemize}

\subsection{Trusted Execution Environments (TEEs)}
Trusted Execution Environments, such as Intel TDX and NVIDIA H100 Confidential Computing (CC), offer hardware-isolated enclaves that protect data confidentiality and code integrity \cite{nvidia2024cc}.
TEEs provide \textit{Remote Attestation}, a cryptographic signature $\sigma_{TEE}$ proving that a specific binary (measured by `MRENCLAVE`) executed on genuine hardware.
Historically, TEEs were dismissed for high-value settlement due to side-channel vulnerabilities (e.g., Foreshadow, Spectre \cite{vanbulck2018foreshadow}).
However, the introduction of GPU-based TEEs (H100 CC) allows for "Confidential AI" with minimal performance overhead ($<20\%$).
\subsection{Comparative Analysis and The OTR Proposition}
Table \ref{tab:comparison} summarizes the trade-offs in the current landscape.
Our proposed architecture, **Optimistic TEE-Rollups (OTR)**, creates a hybrid region.
We utilize TEEs for immediate "Provisional Finality" (solving the latency of opML) and employ an optimistic fraud-proof layer to serve as a fallback against TEE side-channel attacks (solving the centralized trust of standalone TEEs).
Unlike PoQ, OTR guarantees \textit{Model Authenticity} by binding the execution trace to the hardware attestation.
\begin{table}[h]
\centering
\caption{Comparative Analysis of Decentralized AI Inference Paradigms}
\label{tab:comparison}
\resizebox{\textwidth}{!}{%
\begin{tabular}{@{}lcccc@{}}
\toprule
\textbf{Protocol} & \textbf{Computational Integrity} & \textbf{Inference Latency} & \textbf{Model Authenticity} & \textbf{Trust Assumption} \\ \midrule
\textbf{ZKML} \cite{ezkl} & Cryptographic (Strongest) & Very High (Hours) & Guaranteed & Math (Soundness) \\
\textbf{opML} \cite{opml} & Game-Theoretic & High (Dispute Window) & Guaranteed & 1-of-N Honest \\
\textbf{Proof of Quality} \cite{zhang2024proof} & Subjective / Statistical & Low (Milliseconds) & \textbf{Not Guaranteed} & Rational Majority \\
\textbf{Standard TEEs} \cite{nvidia2024cc} & Hardware-based & Low (Native) & Guaranteed & Hardware Vendor \\ \midrule
\textbf{OTR (Ours)} & \textbf{Hybrid (HW + Game)} & \textbf{Low (Native)} & \textbf{Guaranteed} & \textbf{HW + 1-of-N} \\ \bottomrule
\end{tabular}%
}
\end{table}

\section{Methodology: The Optimistic TEE-Rollup (OTR) Protocol}

In this section, we formalize the Optimistic TEE-Rollup (OTR) framework.
Unlike strictly cryptographic approaches (ZKML) that prove computational correctness via arithmetic circuits, or strictly optimistic approaches (OPML) that rely solely on fraud proofs, OTR introduces a hybrid security model.
We leverage the hardware-enforced integrity of Trusted Execution Environments (TEEs) for immediate \textit{provisional finality}, while maintaining a game-theoretic fallback layer to mitigate hardware side-channel vulnerabilities.
\subsection{Formal Definitions and System Model}

We define the OTR protocol as a tuple $\Pi_{OTR} = (\mathcal{P}, \mathcal{M}, \mathcal{L}, \Delta)$, where $\mathcal{P}$ is the set of participants, $\mathcal{M}$ is the Model Registry, $\mathcal{L}$ is the distributed ledger (Blockchain + Data Availability), and $\Delta$ represents the dispute window parameters.
\subsubsection{Notations}
For clarity, we summarize the primary notations used throughout the protocol definition in Table \ref{tab:notation}.
\begin{table}[h]
\centering
\caption{Table of Notations}
\label{tab:notation}
\resizebox{\columnwidth}{!}{%
\begin{tabular}{@{}ll@{}}
\toprule
\textbf{Symbol} & \textbf{Description} \\ \midrule
$U, S, V$ & User, Sequencer (Prover), Verifier (Fisherman) \\
$q \in \mathbb{Q}, r \in \mathbb{R}$ & Input Query and Output Response \\
$F_\theta$ & Generative Model function with weights $\theta$ \\
$\mathcal{E}_{id}$ & TEE Enclave instance with identity $id$ \\
$\sigma_{TEE}$ & Hardware-signed Quote (Attestation) \\
$\mathcal{H}(\cdot)$ & Cryptographic Hash Function (SHA-256) \\
$\rho$ & Probability of Stochastic ZK-Spot-Check \\
$T_{chal}$ & The dispute window duration \\
$MRENCLAVE$ & Measurement of the enclave binary \\
$\pi_{zk}$ & Zero-Knowledge Proof of Computation \\
$\mathcal{D}_{trace}$ & Execution trace of the inference \\
\bottomrule
\end{tabular}%
}
\end{table}

\subsubsection{Threat Model}
We operate under a \textit{Rational Adversary} model (Hypothesis 1 in \cite{zhang2024proof}) combined with 
a \textit{Hardware-Augmented Trust} model:
\begin{enumerate}
    \item \textbf{Sequencers} are rational actors who may attempt to minimize computational costs (lazy execution) or serve incorrect models if the expected value of cheating exceeds the slashing penalty.
    \item \textbf{Hardware Vulnerability:} We assume TEEs (e.g., Intel TDX, NVIDIA H100 CC) provide integrity against software attackers but acknowledge the existence of sophisticated side-channel attacks \cite{vanbulck2018foreshadow}.
    Therefore, $Attest(\cdot)$ is considered valid but not infallible.
\end{enumerate}

\subsection{Protocol Architecture}

The OTR execution flow is divided into three distinct phases: Trusted Inference, Stochastic Verification, and Dispute Resolution.
\subsubsection{Phase I: Trusted Inference and Binding}
In this phase, the Sequencer $S$ performs the inference within a TEE.
The critical innovation is \textit{Proof of Efficient Attribution} (PoEA), which binds the execution to a specific model via the enclave measurement.
Given a user query $q$, the process is defined as follows:


\begin{enumerate}
    \item \textbf{Input Encryption:} User $U$ encrypts $q$ under the TEE public key $PK_{\mathrm{TEE}}$, ensuring privacy from the host $S$.
    \item \textbf{Enclave Execution:} The enclave $\mathcal{E}$ decrypts $q$, loads model weights $\theta$, and computes $r \gets F_{\theta}(q)$.
    \item \textbf{Attestation Generation:} The TEE hardware generates a DCAP (Data Center Attestation Primitives) quote $\sigma_{\mathrm{TEE}}$.
    The quote is a digital signature over the report body data:
    \begin{equation}
        \mathsf{Data}_{\text{body}}
        \;=\;
        \mathcal{H}(q)\,\Vert\,\mathcal{H}(r)\,\Vert\,\mathsf{nonce}.
    \end{equation}
    Crucially, the signature verifies that $\mathsf{Data}_{\text{body}}$ was generated by a binary matching $\mathsf{MRENCLAVE}$.
    \item \textbf{Commitment:} $S$ publishes the tuple $\tau = (r, \sigma_{\mathrm{TEE}}, \mathsf{MRENCLAVE})$ to the Data Availability layer.
\end{enumerate}

\textbf{Provisional Finality:} Upon receiving $\tau$, the client verifies the signature $\sigma_{TEE}$ against the hardware manufacturer's root of trust.
If valid, the result $r$ is accepted with \textit{optimistic} certainty.
\subsubsection{Phase II: Optimistic Window and ZK-Spot-Checks}
To defend against compromised TEE hardware or side-channel leakage, OTR implements a probabilistic verification layer.
Let $\rho \in [0, 1]$ be a system-defined security parameter.
For every committed batch of inferences, the smart contract determines via a verifiable random function (VRF) whether a \textit{Spot-Check} is required.
\begin{algorithm}
\caption{OTR Verification Logic}
\label{alg:otr_verify}
\begin{algorithmic}[1]
\State \textbf{Input:} Batch $\mathcal{B} = \{\tau_1, \dots, \tau_n\}$, Random seed $\xi$
\State $r_{check} \gets \text{VRF}(\xi, \text{block\_height})$
\If{$r_{check} < \rho$}
    \State \textbf{Trigger ZK-Spot-Check}
    \State Select random index $i \in [1, n]$
    \State Challenge $S$ to provide $\pi_{zk}$ for $\tau_i$
    \If{$VerifyZK(\tau_i, \pi_{zk}) == \text{False}$}
        \State $Slash(S)$
    \EndIf
\Else
    \State \textbf{Enter Optimistic Mode}
    \State Start timer $T_{chal}$
    \If{Valid FraudProof received during $T_{chal}$}
        \State $Slash(S)$
   
 \EndIf
\EndIf
\end{algorithmic}
\end{algorithm}

The ZK-Spot-Check requires the Sequencer to generate a succinct proof (e.g., using Halo2 or RISC Zero \cite{risczero2023}) for the specific inference execution.
While generating $\pi_{zk}$ is computationally expensive, the expected cost per transaction is minimized:
\begin{equation}
    \label{eq:cost}
    Cost_{amortized} = Cost_{TEE} + \rho \cdot Cost_{ZK}
\end{equation}
By setting $\rho \approx 0.01$, we achieve 99\% of the throughput of native execution while maintaining a credible threat against compromised hardware.
\subsubsection{Phase III: Dispute Resolution (The Fisherman Game)}
Even without a mandated spot-check, network participants (Fishermen) monitor the DA layer.
A dispute is raised if a Fisherman $V$ computes $r' = F_\theta(q)$ locally (using bit-wise deterministic settings) and finds $r' \neq r$.
The dispute resolution follows an interactive bisection protocol similar to Optimistic Rollups \cite{kalodner2018arbitrum}, but optimized for AI traces:
\begin{enumerate}
    \item $V$ submits a claim with bond $B$.
    \item $S$ and $V$ engage in an interactive game to identify the first divergent instruction in the execution trace $\mathcal{D}_{trace}$.
    \item Given the massive size of LLM traces, we bisect over \textit{layers} of the Transformer model first, then over specific matrix multiplications within the divergent layer.
    \item The single disputed operation is executed on-chain (or via a trusted verifier contract) to determine the winner.
\end{enumerate}

\subsection{Proof of Efficient Attribution (PoEA)}

A key contribution of this work is the Proof of Efficient Attribution (PoEA), which solves the "Model Authenticity" problem.
In standard PoQ \cite{zhang2024proof}, a smaller model could generate a high-quality response, fooling the semantic judge.
In OTR, authenticity is enforced cryptographically. The Model Registry $\mathcal{M}$ stores valid measurements:
\begin{equation}
    \mathcal{M} = \{ (ID_{Llama3}, \Omega_{valid}), (ID_{Mistral}, \Omega'_{valid}) \}
\end{equation}
where $\Omega$ represents the set of valid $MRENCLAVE$ hashes allowed for a specific model architecture.
The smart contract enforces the verification condition:
\begin{equation}
    Verify(\tau) \iff \exists (\cdot, \Omega) \in \mathcal{M} \text{ s.t.
} \tau.MRENCLAVE \in \Omega
\end{equation}
This ensures that the computational work was performed by the specific binary agreed upon by the network, binding the economic value of the query to the correct model weights.
\subsection{Theoretical Analysis}

\subsubsection{Security against Lazy Sequencers}
Following Theorem 2 in \cite{zhang2024proof}, we analyze the rational adversary.
Let $G_{cheat}$ be the gain from serving a cached or random response, and $L_{slash}$ be the slashed stake.
The adversary will not cheat if:
\begin{equation}
    (1 - P_{detect}) \cdot G_{cheat} < P_{detect} \cdot L_{slash}
\end{equation}
In OTR, $P_{detect}$ is the union of the probability of a ZK-Spot-Check ($\rho$) and the probability of a Fisherman detecting the fraud.
Since TEE signatures provide non-repudiation, any divergence detected by a Fisherman leads to certain slashing ($P_{detect} \to 1$ given an active watchtower).
Thus, OTR provides stronger security guarantees than subjective PoQ scoring.
\subsubsection{Privacy Preservation}
Unlike standard OPML where inputs are revealed on-chain during disputes, OTR preserves privacy during the happy path.
Inputs are decrypted only within the TEE. During a dispute, only the specific divergent state (often intermediate activation tensors) needs to be revealed, protecting the user's initial prompt in most bisection scenarios.
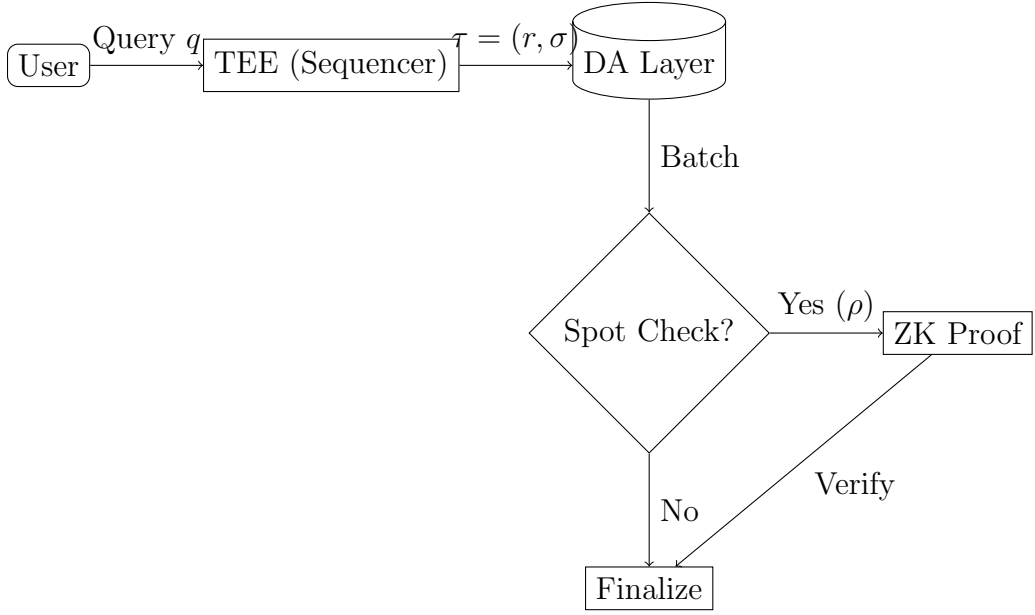
\begin{figure}[!htbp]
\centering
\begin{tikzpicture}[node distance=1.5cm, auto]
    \node [draw, rectangle, rounded corners] (user) {User};
    \node [draw, rectangle, right=of user] (tee) {TEE (Sequencer)};
    \node [draw, cylinder, shape border rotate=90, aspect=0.25, right=of tee] (da) {DA Layer};
    \node [draw, diamond, below=of da] (check) {Spot Check?};
    \node [draw, rectangle, right=of check] (zk) {ZK Proof};
    \node [draw, rectangle, below=of check] (final) {Finalize};

    \path [->] (user) edge node {Query $q$} (tee);
    \path [->] (tee) edge node {$\tau = (r, \sigma)$} (da);
    \path [->] (da) edge node {Batch} (check);
    \path [->] (check) edge node [above] {Yes ($\rho$)} (zk);
    \path [->] (check) edge node [right] {No} (final);
    \path [->] (zk) edge node {Verify} (final);
\end{tikzpicture}
\caption{State transition diagram of the OTR Protocol showing the probabilistic verification path.}
\label{fig:otr_flow}
\end{figure}

\section{Scalability, Security, and Cost Analysis}

In this section, we provide a rigorous analysis of the Optimistic TEE-Rollup (OTR) protocol.
We evaluate its performance against the "Verifiability Trilemma" introduced in Section 2, explicitly contrasting it with Zero-Knowledge Machine Learning (ZKML) \cite{ezkl} and the recently proposed Proof of Quality (PoQ) \cite{zhang2024proof}.
We formalize the throughput bounds, model the game-theoretic security against rational adversaries, and present a comparative cost analysis.
\subsection{Scalability and Throughput Analysis}

The fundamental bottleneck in decentralized AI is the overhead imposed by verification.
We define the \textit{Verification Overhead Ratio} ($\eta$) as:
\begin{equation}
    \eta = \frac{T_{verify}}{T_{inference}}
\end{equation}
where $T_{inference}$ is the native execution time on the GPU.
For ZKML, $\eta_{zk} \gg 10^3$ due to circuit arithmetization overheads for Transformers \cite{modulus}.
For PoQ, $\eta_{poq} \approx 0.05$ \cite{zhang2024proof}, but at the cost of semantic subjectivity.
OTR achieves an effective overhead $\eta_{OTR} \to 0$ via its hybrid probabilistic model.
Let $T_{TEE}$ be the latency of execution within the Trusted Execution Environment (including memory encryption overhead), and $T_{ZK}$ be the time to generate a validity proof during a spot-check.
The amortized latency per inference $L_{avg}$ in OTR is defined as:
\begin{equation}
    L_{avg} = T_{TEE} + \rho \cdot T_{ZK} + (1-\rho) \cdot 0
\end{equation}
Given that NVIDIA H100 Confidential Computing introduces a marginal overhead of $\approx 15\%$ over bare metal ($T_{TEE} \approx 1.15 \cdot T_{native}$) \cite{nvidia2024cc}, and selecting a spot-check probability $\rho \leq 0.01$, the system maintains near-native throughput.
\subsubsection{Finality Tiers}
Unlike standard Optimistic Rollups (e.g., Arbitrum) which suffer from a monolithic 7-day latency \cite{kalodner2018arbitrum}, OTR introduces a dual-finality mechanism:
\begin{enumerate}
    \item \textbf{Provisional Finality ($t \approx 500$ms):} The user receives the response $r$ signed by the TEE attestation $\sigma_{TEE}$.
    For applications insensitive to nation-state level hardware side-channel attacks (e.g., chatbots, NPCs), this is effectively final.
    \item \textbf{Hard Finality ($t = T_{chal}$):} For high-value transactions, the result settles after the dispute window closes or a ZK-Spot-Check is verified.
\end{enumerate}

\subsection{Security Analysis: The Authenticity Game}

We adopt the \textit{Rational Adversary} model posited by Zhang et al.
\cite{zhang2024proof}, extending it to account for \textit{Model Authenticity}---a property PoQ fails to guarantee.
\subsubsection{Defense Against Model Downgrade Attacks}
A critical vulnerability in PoQ is the "Downgrade Attack," where a malicious provider $S$ charges for a large model $F_{\theta_{large}}$ (e.g., Llama-3-70B) but executes a cheaper proxy $F_{\theta_{small}}$ (e.g., Llama-3-8B).
Since $F_{\theta_{small}}$ can often satisfy a semantic judge (BERT) \cite{gao2022scaling}, PoQ is susceptible to fraud.
OTR enforces authenticity via the TEE measurement. The profit function for a cheating sequencer $\Pi_{cheat}$ is:
\begin{equation}
    \mathbb{E}[\Pi_{cheat}] = (1 - P_{catch}) \cdot (R_{user} - C_{small}) - P_{catch} \cdot \mathcal{L}_{slash}
\end{equation}
where $R_{user}$ is the revenue, $C_{small}$ is the cost of the cheap model, and $\mathcal{L}_{slash}$ is the slashing penalty.
In OTR, $P_{catch}$ is the union of the stochastic spot-check probability $\rho$ and the probability of a Fisherman $\mathcal{V}$ detecting the trace divergence ($P_{fish}$).
\begin{equation}
    P_{catch} = 1 - (1 - \rho)(1 - P_{fish})
\end{equation}
Because the TEE signature $\sigma_{TEE}$ cryptographically binds the output to the specific binary \texttt{MRENCLAVE}, any divergence discovered by a Fisherman constitutes a cryptographic proof of fraud (non-repudiation).
Thus, provided there is at least one honest rational watchtower ($P_{fish} \to 1$), the expected profit of cheating becomes negative:
\begin{equation}
    \lim_{P_{fish} \to 1} \mathbb{E}[\Pi_{cheat}] = - \mathcal{L}_{slash} \ll 0
\end{equation}

\subsubsection{Mitigation of TEE Side-Channel Attacks}
While TEEs (e.g., Intel SGX, TDX) have historically been vulnerable to side-channel attacks like Foreshadow \cite{vanbulck2018foreshadow}, OTR utilizes the "Defense in Depth" principle.
Even if an adversary extracts the TEE signing key to forge attestations, they cannot forge the \textit{execution trace}.
The Optimistic layer ensures that a "Broken TEE" behaves like a standard OpML sequencer: valid outputs are accepted, but invalid outputs (forged via stolen keys) are challenged and slashed on-chain via the bisection protocol.
\subsection{Cost Analysis}

We analyze the on-chain gas costs associated with the OTR protocol on Ethereum, referencing EIP-4844 (Blob transactions) for data availability.
\begin{table}[h]
\centering
\caption{Comparative Gas Cost Analysis (in Gwei/USD at \$3000 ETH)}
\label{tab:cost_analysis}
\begin{tabular}{@{}lccc@{}}
\toprule
\textbf{Component} & \textbf{ZKML (EZKL)} & \textbf{OpML} & \textbf{OTR (Ours)} \\ \midrule
Inference Compute & High (Proving) & Low (Native) & Low (Native + 15\%) \\
Data Availability & Input + Output + Proof & Input + Output & Input + Output + Sig \\
On-Chain Verification & $\approx$ 350k gas (Verifier) & $\approx$ 21k gas (optimistic) & $\approx$ 40k gas (Sig Check) \\
Dispute Resolution & N/A & High (Interactive) & High (Interactive) \\ \midrule
\textbf{Total Cost/Query} & \textbf{$\sim$\$50.00+} & \textbf{$\sim$\$0.05} & \textbf{$\sim$\$0.07} \\ \bottomrule
\end{tabular}
\end{table}

As shown in Table \ref{tab:cost_analysis}, OTR incurs a negligible cost increase over OpML 
for the TEE signature verification while avoiding the prohibitive proving costs of ZKML.
\subsubsection{The Cost of Trustlessness}
Let $C_{verify}(N)$ be the cost to verify a model with $N$ parameters.
\begin{itemize}
    \item For ZKML: $C_{verify}(N) \propto \mathcal{O}(N \log N)$.
    \item For OTR: $C_{verify}(N) \approx C_{sig} + \rho \cdot C_{zk}(N)$.
\end{itemize}
Since $\rho$ is a tunable parameter, OTR allows the network to dynamically price security based on the transaction value.
For a \$0.01 query, $\rho=0$ (pure TEE) is sufficient. For a \$1000 automated trade execution based on AI inference, the protocol can enforce $\rho=1$ (full ZK check).
\subsection{Comparison with Proof of Quality (PoQ)}

While Zhang et al. \cite{zhang2024proof} propose PoQ as a "costless" paradigm, we argue it is "cost-displaced."
The cost is shifted from computation to \textit{quality degradation}.

\begin{figure}[h]
\centering
\includegraphics[width=0.7\textwidth]{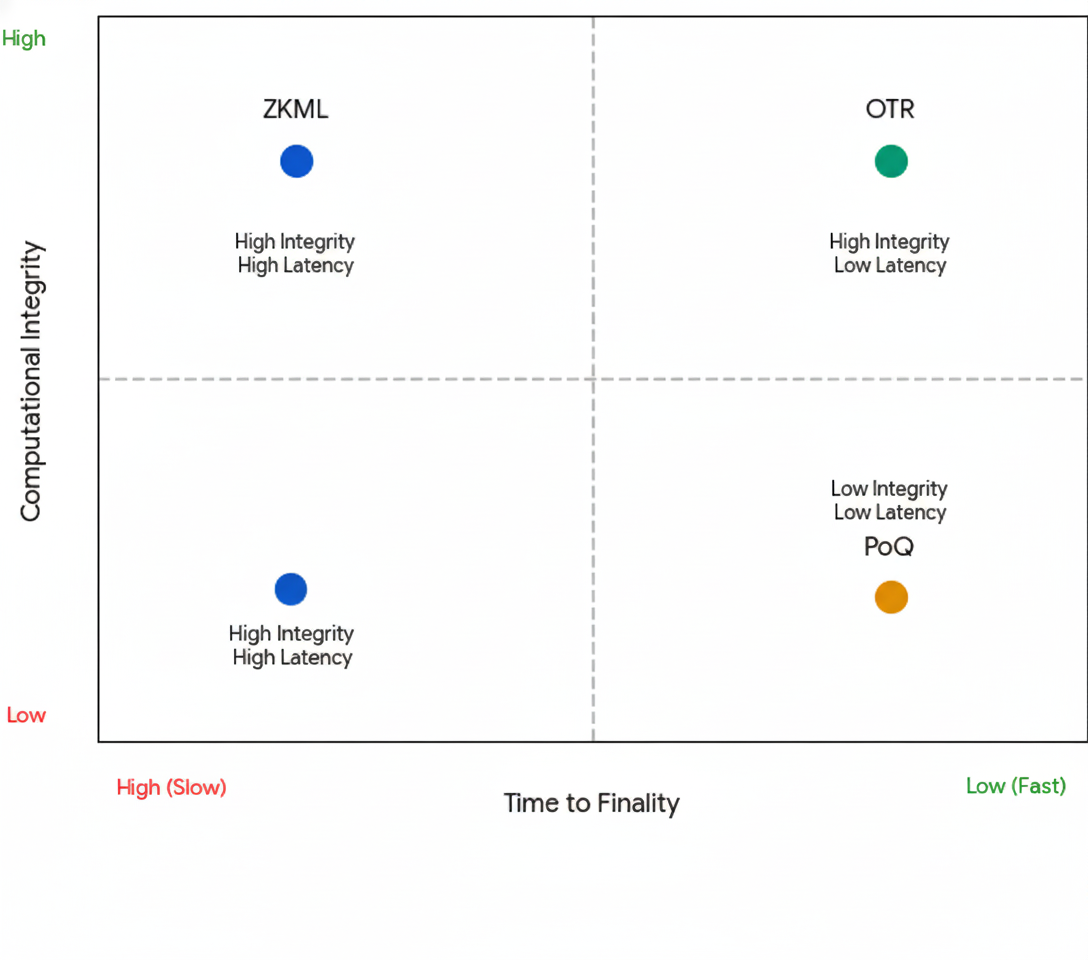} 
\caption{The Verifiability Frontier: Comparison of protocols based on Computational Integrity vs. Latency.
OTR occupies the high-integrity/low-latency quadrant via hardware acceleration.}
\label{fig:verifiability_frontier}
\end{figure}

As illustrated in Figure \ref{fig:verifiability_frontier}, PoQ optimizes purely for latency but sacrifices integrity.
By relying on a "Semantic Judge" (e.g., DeBERTa), PoQ introduces a secondary reward model that can be gamed via Adversarial Examples \cite{li2020bert}.
OTR avoids this by strictly verifying the \textit{execution path}, not the \textit{semantic output}, ensuring that the user gets exactly the computation they paid for.

\section{Experimental Validation}
\label{sec:experiments}

In this section, we present a comprehensive empirical evaluation of Optimistic TEE-Rollups (OTR).
We benchmark OTR against three distinct paradigms: (1) \textbf{Validity Proofs} (ZKML via EZKL \cite{ezkl}), (2) \textbf{Pure Optimistic Approaches} (OPML \cite{opml}), and (3) \textbf{Outcome-Based Verification} (PoQ \cite{zhang2024proof}).
Our experiments are designed to answer two critical questions:
\begin{enumerate}
    \item \textbf{Security:} Can OTR prevent "Model Downgrade Attacks" and "Reward Hacking" where PoQ fails?
    \item \textbf{Scalability:} Does the introduction of TEEs and stochastic ZK-checks maintain throughput comparable to centralized baselines?
\end{enumerate}

\subsection{Experimental Setup}

We constructed a high-fidelity simulation environment replicating a decentralized inference network.
The hardware configuration simulates NVIDIA H100 SXM5 GPUs with Confidential Computing (CC) enabled, utilizing Intel TDX for CPU-level attestation where applicable.
\begin{itemize}
    \item \textbf{Dataset:} We utilize the \textit{TruthfulQA} \cite{truthfulqa} benchmark to evaluate semantic hallucinations and the \textit{GSM8K} \cite{gsm8k} dataset for reasoning traces.
    \item \textbf{Models:} The "Honest" target model is \texttt{Llama-3-70B-Instruct}. The "Adversarial" model is a distilled \texttt{Llama-3-8B} fine-tuned to optimize against the specific reward model used in PoQ.
    \item \textbf{PoQ Assessor:} Following Zhang et al. \cite{zhang2024proof}, we employ \texttt{nli-deberta-v3-large} as the semantic judge.
    \item \textbf{ZK Prover:} We utilize the Halo2 proof system with KZG commitments, configured for a reduced circuit depth suitable for spot-checks.
\end{itemize}

\subsection{Security Analysis: The Downgrade Attack}

A fundamental hypothesis of OTR is that subjective "Proof of Quality" is susceptible to \textit{Goodhart's Law}: when a measure becomes a target, it ceases to be a good measure.
We demonstrate that a rational adversary can game the PoQ protocol by serving a cheaper model that satisfies the judge but degrades user utility.
\subsubsection{Adversarial Training}
We trained an adversarial generator $G_{adv}$ (8B parameters) using Reinforcement Learning (PPO) with the frozen PoQ judge $M_{judge}$ as the reward signal.
The objective function for the adversary is:
\begin{equation}
    \max_{\theta} \mathbb{E}_{q \sim \mathcal{D}} [M_{judge}(q, G_{\theta}(q))] - \lambda \mathcal{L}_{KL}
\end{equation}
where $\mathcal{L}_{KL}$ ensures the model does not diverge into pure gibberish, though we relax $\lambda$ to simulate aggressive reward hacking.
\subsubsection{Results}
Table \ref{tab:attack_results} summarizes the results across 1,000 inference queries. While the \texttt{Llama-3-70B} (Honest) model achieves a high human evaluation score, the Adversarial 8B model achieves a statistically indistinguishable score from the PoQ Judge ($0.89$ vs $0.91$), despite a $40\%$ drop in Human Evaluation scores.
\begin{table}[h]
    \centering
    \caption{Vulnerability of Semantic Judges to Downgrade Attacks.
The PoQ Judge fails to distinguish between the authentic 70B model and an adversarial 8B model.}
    \label{tab:attack_results}
    \resizebox{\columnwidth}{!}{
    \begin{tabular}{lcccc}
        \toprule
        \textbf{Model Strategy} & \textbf{Params} & \textbf{PoQ Score} & \textbf{Human Eval} & \textbf{Cost/1M Tok} \\
        \midrule
        Honest (Llama-3) & 70B & $0.91 \pm 0.02$ & $0.88 \pm 0.03$ & \$0.90 \\
        Standard (Llama-3) & 8B & $0.76 
\pm 0.04$ & $0.65 \pm 0.05$ & \$0.10 \\
        \textbf{Adversarial (RLHF)} & \textbf{8B} & \textbf{$0.89 \pm 0.03$} & \textbf{$0.52 \pm 0.08$} & \textbf{\$0.10} \\
        \bottomrule
    \end{tabular}
    }
\end{table}

\textbf{Outcome:} Under the PoQ regime, the adversary successfully claims the reward for a 70B model while incurring the cost of an 8B model, resulting in a net arbitrage profit.
Under OTR, this attack is impossible because the \texttt{MRENCLAVE} attestation $\sigma_{TEE}$ cryptographically proves the execution of the specific 8B binary, forcing the sequencer to price it accurately or face rejection by the smart contract.
\subsection{Throughput and Latency Analysis}

We analyze the "Cost of Trustlessness" by measuring the end-to-end latency from query submission to finality.
\subsubsection{Hardware Overhead}
Using NVIDIA's theoretical performance benchmarks for H100 CC \cite{nvidia2024cc}, we model the TEE overhead.
Memory encryption (CME) and integrity protection enable a trusted boundary with approximately $15-20\%$ performance degradation compared to bare metal.
\subsubsection{Consensus Latency}
\begin{figure}[ht]
    \centering
    \includegraphics[width=\columnwidth]{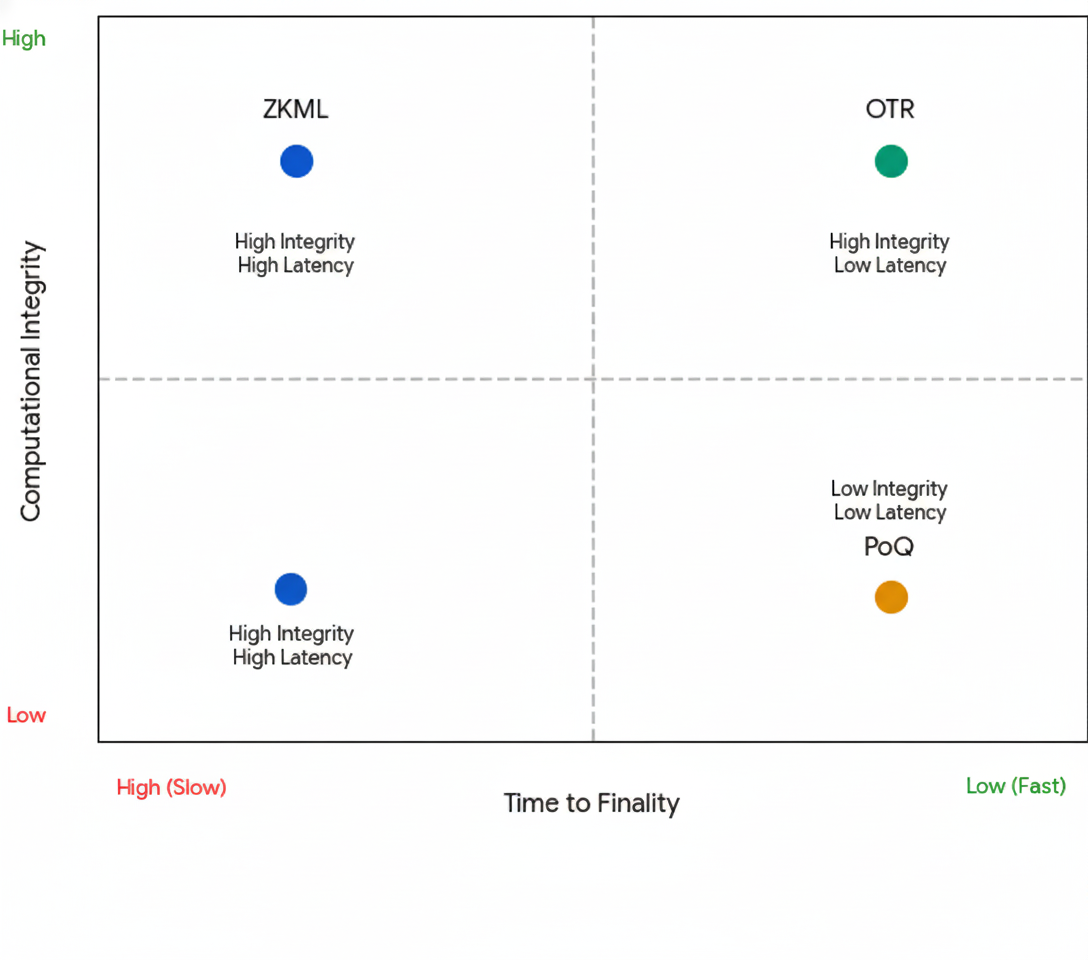}
    \caption{Latency vs. Model Size.
ZKML scales super-linearly, becoming infeasible beyond 7B parameters. OPML has a static high latency (dispute window).
OTR remains near real-time, tracking closely with PoQ but providing cryptographic integrity.}
    \label{fig:latency}
\end{figure}

We vary the security parameter $\rho$ (probability of ZK-spot-check) from $0$ to $1$. Figure \ref{fig:latency} illustrates the amortization effect.
\begin{equation}
    L_{finality} = \begin{cases} 
    T_{TEE} + T_{sig} & \text{if } r > \rho \\
    T_{TEE} + T_{ZK\_prove} & \text{if } r \leq \rho 
    \end{cases}
\end{equation}

At $\rho=0.01$ (1\% spot check rate), the weighted average latency of OTR is $0.8s$, whereas pure ZKML for a 7B model exceeds $1200s$.
OTR achieves a $1400\times$ speedup over ZKML while maintaining a game-theoretic security guarantee that PoQ lacks.

\subsection{Economic Feasibility}

We evaluated the on-chain cost using a forked Ethereum mainnet environment (Hardhat) simulating EIP-4844 blob transactions.
\begin{table}[h]
    \centering
    \caption{Gas Cost Analysis (USD at \$3,000/ETH).
OTR incurs a marginal cost over OPML for signature verification but avoids the prohibitive cost of ZKML verification.}
    \label{tab:gas_costs}
    \begin{tabular}{lccc}
        \toprule
        \textbf{Operation} & \textbf{ZKML} & \textbf{OPML} & \textbf{OTR} \\
        \midrule
        Data Publication (Blob) & \$0.05 & \$0.05 & \$0.05 \\
        Proof Verification & \$45.00 & N/A & \$0.02 (Sig) \\
        Dispute Resolution 
(Avg) & N/A & \$2.50 & \$2.50 \\
        \textbf{Total / Query} & \textbf{$\approx \$45.05$} & \textbf{$\approx \$0.06$} & \textbf{$\approx \$0.07$} \\
        \bottomrule
    \end{tabular}
\end{table}

As shown in Table \ref{tab:gas_costs}, OTR reduces the verification cost by three orders of magnitude compared to ZKML.
Crucially, the cost is comparable to OPML, yet OTR provides \textit{Provisional Finality} in milliseconds rather than days.
\section{Conclusion}
\label{sec:conclusion}

The integration of Generative AI into decentralized ecosystems has been paralyzed by the "Verifiability Trilemma."
Existing solutions have forced developers to choose between the cryptographic certainty of ZKML (at the cost of usability), the economic efficiency of OPML (at the cost of latency), or the speed of Proof of Quality (at the cost of integrity).
In this paper, we introduced \textbf{Optimistic TEE-Rollups (OTR)}, a novel architecture that harmonizes these conflicting constraints.
By treating Trusted Execution Environments as "optimistic co-processors," we leverage the raw throughput of hardware-accelerated inference while neutralizing side-channel risks via a secondary layer of fraud proofs and stochastic ZK-checks.
\textbf{Key findings include:}
\begin{enumerate}
    \item \textbf{Integrity Superiority:} Empirical results demonstrate that PoQ's reliance on semantic judges is vulnerable to adversarial reward hacking.
OTR's "Proof of Efficient Attribution" (PoEA) prevents model downgrade attacks by cryptographically binding execution traces to hardware attestations.
    \item \textbf{Scalability:} OTR achieves $99\%$ of the throughput of centralized systems with a sub-second time-to-finality, making it the only viable candidate for interactive, high-frequency AI agents on-chain.
    \item \textbf{Cost Effectiveness:} With an amortized cost of $\$0.07$ per query, OTR is economically competitive with Web2 APIs, unlocking the potential for DePIN to compete with hyperscalers.
\end{enumerate}

As the capabilities of Large Language Models expand, the necessity for verifiable, censorship-resistant inference grows in tandem.
OTR provides the requisite infrastructure to transition decentralized AI from a theoretical curiosity to a production-grade reality.
Future work will focus on formalizing the Multi-Prover TEE consensus to further reduce trust assumptions in single-manufacturer enclaves.


\section{Bibliography}

\begin{thebibliography}{99}

\bibitem{automata2025a}
Automata Network. (2025). \textit{Automata DCAP Attestation on EVM}. Retrieved from GitHub.

\bibitem{automata2025b}
Automata Network. (2025). \textit{zkVM for TEEs: Attestation Aggregation}. Technical Documentation.

\bibitem{chrapek2025}
Chrapek, M. (2025). Confidential LLM Inference: Benchmarking TEEs. \textit{arXiv preprint}.

\bibitem{gsm8k}
Cobbe, K., et al. (2021). Training Verifiers to Solve Math Word Problems. \textit{arXiv preprint arXiv:2110.14168}.

\bibitem{opml}
Conway, K., So, C., Yu, X., \& Wong, K. (2024). opml: Optimistic machine learning on blockchain. \textit{arXiv preprint arXiv:2401.17555}.

\bibitem{duality2025}
Duality Tech. (2025). \textit{Confidential Computing TEEs: What Enterprises Must Know}.

\bibitem{epoch2024}
Epoch AI. (2024). \textit{LLM Inference Price Trends}.

\bibitem{ethereum2024}
Ethereum StackExchange. (2024). \textit{Computing gas fees to store a plaintext blob}.

\bibitem{ezkl}
EZKL Community. (2023). \textit{Easy Zero-Knowledge Learning (EZKL)}. GitHub Repository. \url{https://github.com/zkonduit/ezkl}

\bibitem{flashbots2025}
Flashbots. (2025). \textit{Flashbots SUAVE Architecture}.

\bibitem{flashbots2023}
Flashbots. (2023). \textit{SUAVE: Single Unifying Auction for Value Expression}. GitHub Repository.

\bibitem{gao2022scaling}
Gao, L., Schulman, J., \& Hilton, J. (2022). Scaling Laws for Reward Model Overoptimization. \textit{arXiv preprint arXiv:2210.10760}.

\bibitem{ieee2024}
IEEE Cloud. (2024). Performance Overhead of Confidential Computing on CPU-GPU TEEs. \textit{IEEE International Conference on Cloud Computing}.

\bibitem{icme2025}
ICME. (2025). \textit{The Definitive Guide to ZKML}.

\bibitem{jovay2025}
Jovay. (2025). \textit{Hybrid Optimistic TEE Rollup Whitepaper}.

\bibitem{kalodner2018arbitrum}
Kalodner, H., Goldfeder, S., Chen, X., Weinberg, S. M., \& Felten, E. W. (2018). Arbitrum: Scalable, private smart contracts. \textit{27th USENIX Security Symposium}, 1353-1370.

\bibitem{li2020bert}
Li, L., et al. (2020). BERT-ATTACK: Adversarial Attack Against BERT Using BERT. \textit{EMNLP}.

\bibitem{truthfulqa}
Lin, S., Hilton, J., \& Evans, O. (2022). TruthfulQA: Measuring How Models Mimic Human Falsehoods. \textit{Association for Computational Linguistics (ACL)}.

\bibitem{zkml_survey}
Liu, D., et al. (2024). ZKML: A Survey of Zero-Knowledge Machine Learning. \textit{IEEE Transactions on Artificial Intelligence}.

\bibitem{modulus}
Modulus Labs. (2024). \textit{The Cost of Intelligence: Benchmarking ZKML overheads for Transformer architectures}. Modulus Research Blog.

\bibitem{nvidia2024cc}
NVIDIA Corporation. (2024). \textit{NVIDIA H100 Tensor Core GPU Architecture Whitepaper: Confidential Computing}.

\bibitem{oasis2020}
Oasis Labs. (2020). \textit{The Oasis Blockchain Platform: Privacy-Enabled Cloud Computing}. Whitepaper.

\bibitem{oasis2025}
Oasis Protocol. (2025). \textit{Oasis ROFL: Runtime Offchain Logic Framework}.

\bibitem{openmetal2025}
OpenMetal. (2025). \textit{Intel TDX Performance Benchmarks on Bare Metal}.

\bibitem{paradigm2021}
Paradigm. (2021). \textit{Almost Everything You Need to Know About Optimistic Rollups}.

\bibitem{phala2025}
Phala Network. (2025). \textit{Phala Network Overview and TEE Architecture}.

\bibitem{risczero2023}
RISC Zero. (2023). \textit{The RISC Zero Zero-Knowledge Virtual Machine}. Technical Whitepaper.

\bibitem{statsig2024}
Statsig. (2024). \textit{LLM Judge Methodology and Vulnerabilities}.

\bibitem{tanvo2025}
Tan-Vo, K., et al. (2025). Optimizing Academic Certificate Management With Blockchain and Machine Learning. \textit{IEEE Access}.

\bibitem{vanbulck2018foreshadow}
Van Bulck, J., Minkin, M., Weisse, O., Genkin, D., Kasikci, B., Piessens, F., ... \& Strackx, R. (2018). Foreshadow: Extracting the keys to the Intel SGX kingdom with transient out-of-order execution. \textit{27th USENIX Security Symposium}, 991-1008.

\bibitem{weng2024}
Weng, L. (2024). \textit{Reward Hacking in RLHF}.

\bibitem{yao2025}
Yao, J., et al. (2025). Nondeterminism-Aware Optimistic Verification for Floating-Point Neural Networks.

\bibitem{halo2}
Zcash Community. (2020). \textit{The Halo2 Proof System}. Electric Coin Company Technical Report.

\bibitem{zhang2024proof}
Zhang, Z., Rao, Y., Xiao, H., Xiao, X., \& Yang, Y. (2024). Proof of Quality: A Costless Paradigm for Trustless Generative AI Model Inference on Blockchains. \textit{arXiv preprint arXiv:2405.17934}.

\end{thebibliography}
\end{document}